# AI and the Iterable Epistopics of Risk

Andy Crabtree
School of Computer Science
University of Nottingham
United Kingdom
ORCID ID 0000-0001-5553-6767
**Corresponding author:** andy.crabtree@nottingham.ac.uk

Glenn McGarry
School of Computer Science
University of Nottingham
United Kingdom
ORCID ID 0000-0001-9518-3126

Lachlan Urquhart
Edinburgh Law School
University of Edinburgh
United Kingdom
ORCID ID 0000-0001-5144-5024

Funding information: This work was supported by the Engineering and Physical Sciences Research Council grant number EP/V026607/1.

**Abstract.** The risks AI presents to society are broadly understood to be manageable through 'general calculus', i.e., general frameworks designed to enable those involved in the development of AI to apprehend and manage risk, such as AI impact assessments, ethical frameworks, emerging international standards, and regulations. This paper elaborates how risk is apprehended and managed by a regulator, developer and cyber-security expert. It reveals that risk and risk management is dependent on mundane situated practices not encapsulated in general calculus. Situated practice surfaces 'iterable epistopics', revealing how those involved in the development of AI *know* and subsequently respond to risk and uncover major challenges in their work. The ongoing discovery and elaboration of epistopics of risk in AI a) furnishes a potential program of interdisciplinary inquiry, b) provides AI developers with a means of apprehending risk, and c) informs the ongoing evolution of general calculus.

*The proposal puts in place a well-defined risk-based regulatory approach to ensure AI is developed in ways that respect people's rights and earn their trust.* (EU AI Act 2024)

*Understanding and managing the risks of AI systems will help to enhance trustworthiness, and in turn, cultivate public trust.* (NIST 2023)

## 1. INTRODUCTION

Risk management is fundamental to the adoption of AI in society, widely seen and treated as the cornerstone of *trust*, not only by government and regulatory authorities (Keegan 2019) but increasingly by industry (e.g., McKinsey 2019, Deloitte 2020, Gartner 2023, KPMG 2023). The basic premise is that if we understand the risk that AI presents to society, if we make it accountable, then we can take steps to prevent or mitigate risk and this will instil public confidence. Consequently, risk management frameworks for AI proliferate (see, for example, AI Ethicist 2024). A recent survey (Trilateral Research 2021) shows that a broad array of actors in civil society and industry along with international organisations and nation states are engaged in developing risk management frameworks for AI. Risk management frameworks (assessment procedures, ethical principles, regulations, etc.) each provide in their own way a 'calculus' for building trust in AI. Not in the strict mathematical sense, but in the sociological sense of a providing a method or procedure for reasoning about risk and rendering it accountable and thus tractable. For examples see the Ada Lovelace Institute (2020) on algorithmic impact assessments, or the Alan Turing Institute's SUM values (Leslie 2019), or ISO 38507 (2022) standard governing the use of AI in organisations, or the European Union's recent introduction of regulation for AI, the 'AI Act' or AIA (2024). While substantively different, each in their own way nevertheless provides a general calculus for apprehending and managing risks in AI systems. They orient us to particular features of AI that are deemed risky, such as data provenance, quality and representativeness (AIA 2024), so that regardless of particular context and domain we attend to them and take steps to manage them.

Below we reflect on a brief study involving three different actors having different roles in AI development: a systems developer in a medium-sized international medical devices company, a professional hacker or cyber-security expert in a very large international electronics company, and a UK regulator. These actors present a different perspective on risk management as a *mundane preoccupation* in their workaday lives, which is "systematically under-recognised" in the risk society (Wynne 1996). Risk management frameworks are a central feature of the risk society, making the problems of "techno-economic development" (such as the alleged existential threat posed by AI) "tolerable" and reflexively "legitimise" the developments in question (Beck 1986). The introduction of risk management frameworks by industry and state actors alike give "credibility" to political claims that steps are being taken to address the hazards and insecurities of new technologies (Shapin 1995). These frameworks are typically developed by experts – the EU's AI Act (AIA 2024) was heavily influenced by the High Level Ethics Group (HLEG), for example – and ignore the mundane and vernacular: lay knowledge and expertise is treated as "empirically vacuous", of "no real content or authority beyond the parochial, subjective and emotional worlds of its carriers" (Wynne 1996).

While our discussion with actors involved in AI development does not offer general panaceas, it does sensitise us to mundane challenges for AI development that extend beyond their parochial origins. Thus, we find risks within the innovation environment where AI solutions are being developed, in the internal and external operating environment (within the machine and without), and in the regulatory environment. These risks are occasioned by novelty and lack of knowledge, the limitations of AI models and software verification, the impact of complex systems on and the human configuration of AI, and the nature of the regulatory system. Originally, our discussions surface 'iterable epistopics' (Lynch 1993) of risk, i.e., risks that are encountered and thus known to exist in mundane practice which may be further explored and elaborated or iterated through continued investigation in other settings of AI development. Iterable epistopics orient us to and inform our understanding of how risk is apprehended and managed in practice, in contrast to what it looks like from general point of view, and may identify risks not identified by general calculus. The discovery of iterable epistopics of risk may make three potential contributions. It may furnish a potential program of interdisciplinary inquiry oriented to the iteration of risk in AI in mundane practice. It may provide AI developers with a means of apprehending risks occasioned by users and their environment and inform the design of the human machine interface tools required by regulation to prevent risk in the use of AI. It may also contribute to governance, informing general calculus and enabling it to make emerging risks in AI development generally accountable.

## 2. ITERABLE EPISTOPICS?

The idea of 'iterable epistopics' (Lynch 1993) has its origins in and was inspired by ethnomethodology's respecification of the sciences as sciences of practical action, which relocates the achievements of science in local, work-site specific practices rather than general methods (Garfinkel et al. 1981, Garfinkel 2001, Garfinkel 2022). Ethnomethodology (EM) is a branch of sociology. Like other branches of sociology, it has foundational interest in the problem of social order (Garfinkel 1967). Unlike other branches of sociology, EM sees social order as a *constitutive* feature of human action (Korbut 2014), which is to say that order is a built-in feature of everything we say and do and that everything we say and do may therefore be examined for its orderly properties (Sacks 1984). Sociologist Harold Garfinkel explains how he came to coin the term 'ethnomethodology' in the 1950s.

> "Back in 1954 Fred Strodtbeck asked me to listen to the tapes of jurors. The notion occurred to me of analysing their deliberations. They were not acting in their affairs as jurors as if they were scientists. However, they were concerned with such things as adequate accounts, adequate description, and adequate evidence. Thus, you have this interesting acceptance, so to speak, of these magnificent methodological things like 'fact' and 'fancy' and 'opinion' and 'my opinion' and 'your opinion' and 'what we're entitled to say' and 'what the evidence shows' and 'what can be demonstrated' and 'what actually he said' as compared with 'what only you think he said' or 'what he seemed to have said'. Here I am faced with jurors who are doing methodology. It's not a methodology my colleagues would honour if they were attempting to staff the sociology department. [But] that is what ethnomethodology is concerned with. It is an organisational study of a member's knowledge of his ordinary affairs, of his own organised enterprises, where that knowledge is treated by us as part of the same setting that it also makes orderable. (Garfinkel in Hill & Stones Crittenden 1968, edited extract)

EM's treatment of knowledge and methodology moves them from a professional preoccupation in social science research to the preoccupation of ordinary persons in the mundane conduct of their own organised enterprises. This means that topics of knowledge – things like 'fact' and 'fancy', my opinion' and 'your opinion', 'what the evidence shows' and 'what can be demonstrated', etc. – are social objects produced methodologically in situated practices, of juror deliberation for example. Thus, the topics of which knowledge speaks are seen and treated as 'magnificent methodological things', products of the situated practices that order the enterprise in question (Suchman 1987).

As Science and Technology Studies pioneer Mike Lynch puts it, the magnificent methodological things uncovered by EM studies elaborate 'iterable epistopics' (Lynch 1993). Lynch uses the term with respect to standing topics of epistemological discourse: observation, description, replication, testing, measurement, explanation, proof, and so on. Standard topics of knowledge in science. These epistopics are, for Lynch, 'iterable', i.e., they may be subject to further investigation and our understanding of them may thereby be developed further. One might, for example, look at epistopics in different scientific disciplines to see if they are embedded in the same sorts of situated practices (as Lynch demonstrated to critical acclaim, they aren't, which means that scientific knowledge isn't ordered in the same ways in practice and that it therefore makes no sense to speak of a unitary scientific method). One might also cast the eye further afield and investigate what epistopics look like in more mundane affairs, e.g., what measurement looks like on the building site or in a sawmill and other less specialised domains as well. In short, one might look at epistopics in different settings, contexts and domains to 'iterate' or *develop our understanding* of knowledge production in practice. In doing so we may discover other magnificent methodological things, new epistopics uncovered in one domain, not seen before in others, that are germane to understanding the socially organised nature of knowledge production and may feed further iteration.

Lynch's work on scientific practice furnishes us with the notion of iterable epistopics, but it is not necessarily constrained to studies of science. From its infancy EM has had foundational interest in understanding members' knowledge of their own ordinary affairs, be it jury deliberation or a chemistry experiment, and indeed the many and varied organised enterprises ethnomethodologists have studied since 1954 (EM/CA 2021). Epistopics are not the sole prerogative of science, as Lynch demonstrated in investigating epistopics in science *and* ordinary action. The notion of epistopics might usefully be appropriated then to understand risk in AI and how members know and manage risk in their work, whatever their work may be. Iterable epistopics of risk would reveal the magnificent methodological things that risk and risk management turn on in the situated practices of those involved in the development of AI, be they engineers, users, regulators, politicians, etc. Their elaboration would surface what members know of risk as feature of their own organised enterprises, knowledge which makes that enterprise orderable, which in turn shapes AI in society, and which may be iterated and our knowledge be further developed through further investigation.

**3. STUDYING THE ITERABLE EPISTOPICS OF RISK**

Below we describe the practicalities of our studies, including our participants, and address potential methodological concerns with the size and representative nature of our 'sample' and the duration of our study, before moving on to consider our findings and the iterable epistopics of risk uncovered by the study.

**3.1 The study**

The findings reported below are based on discussions with three different parties involved in the development of AI who have direct experience of risk and risk management. These include three members of staff from a manufacturer of medical devices, the company's principal data scientist, project manager, and director of research; a cyber-security expert working on machine learning in vehicles; and a policy expert who worked for the regulator of medical devices in the UK, the MHRA. Why four participants who in one way or another work in the medical sector and another who works on automobiles? There is no strategy in our selection or the juxtaposition of participants. They were partners in a research project focusing on governance and regulation in autonomous systems (TAS 2023), who we know through informal discussion had experience of managing risks in AI and were willing to talk to us candidly about them with an eye towards our being able to publish something of what they said. This may sound trite but in our experience, despite the hype surrounding AI, candid talk about risk while commonplace amongst practitioners is hard to get 'on the record'. Many project partners simply did not want to 'go on the record' and discuss just what they were doing with AI and the risks involved. Even the medical device company required we sign a non-disclosure agreement (NDA) and reports of our discussion had to be approved before use. As one of the reviewers of this paper pointed out, "much discussion of risk is consequentialist." However, reluctance on the part of practitioners to discuss risk as mundane feature of their work means that general calculus is *blind* to many of risks that accompany AI development. There is much to be had from engaging practitioners in the mundane elaboration of risk, as our findings make perspicuous.

From a practical point of view, our discussions with participants could be called interviews or informal interviews or expert interviews, but we think what matters most about them is that we were engaging *with practitioners* who, *as practitioners*, could tell us something of what we were talking about when we talk about the management of risk in AI. One might say we were using an indeterminate notion of 'risk' as a *coat hanger* to drive discussion of participants mundane practices in details of their shop work and shop talk (Garfinkel 2002). The discussions took place via Microsoft Teams. Each lasted around an hour. Two separate discussions were held with the device manufacturer with the same staff present on both occasions, and two with the regulator, both over a 3 month time frame. One meeting took place with the cyber security expert, but documentation was also furnished and they subsequently participated in a project meeting to discuss the study results. Each party discussed risk separately, these were not joint discussions. Our discussions were entirely open-ended. They were not guided by an interview schedule or protocol, there were no pre-set questions, only a declared interest in risk. The discussions were subsequently driven by what our participants wanted to say about their understanding of risk so that we might learn about things we did not know. Each discussion was recorded to permit transcription and subsequent analysis of participants' accounts for their methodological features (Crabtree et al. 2017) and concomitant identification of epistopics of risk.

Five hours of discussion with five participants in three different occupations may sound like an inadequate basis to address the risks of AI. Methodologically, it might appear that our study is too short, involved too few people, and that any findings must be limited in scope as our sample is not representative of the parties involved in managing risks in AI development. These sorts of complaint turn on quantitative reasoning. Ours is a qualitative study. Nonetheless, and as previously articulated in detail by researchers in the interdisciplinary field of Human Computer Interaction (HCI), the duration of a study, representativeness and size of participating cohort does not necessarily (by default) limit its validity or the generalisation of results. *How Many Bloody Examples Do You Want?* (Crabtree et al. 2013), tackles these methodological issues head on, drawing on an 8 hour observational study of one family's 'day out in the country' informing technology for tourism to demonstrate the point. *The Connected Shower* (Crabtree et al. 2020) turns on 4 hours of interview data from 6 households with 12 people and provides generalisable insights into the adoption of personalised Internet of Things services in the home. *Talking in the Library* (Crabtree et al. 1997) is based on 12 minutes of data involving 3 people and provides generalisable insights into the social organisation of search and retrieval informing the design of digital libraries. Xerox's *Unremarkable Computing* paper (Tolmie et al. 2002), which had a significant impact on HCI's understanding of ubiquitous computing, is based on *2 minutes* of data from just *2 people*. Many, many other examples could be provided of short duration studies and/or small amounts of data collected from a small number of participants that produce generalisable results. Further enumeration would be ironic. It's not the numbers that count with qualitative study, but the *analytic insight*. Did we see something new? Did we learn something useful? Do the results have reach?

## 4. STUDY FINDINGS: ITERABLE EPISTOPICS OF RISK

Below our pseudonymous participants David (principal data scientist), Julia (project manager) Rene (director of research), Robert (cyber-security expert) and James (regulator) tell us what they know of risk and risk management as a feature of their own organised enterprises – medical device manufacture, cyber-security evaluation, and regulation of medical devices. Before we explore risk management with them, we first consider their understanding of risk. We note that in either case they draw no distinction between AI and machine learning.

### 4.1 What is risk?

Our brief discussion with practitioners involved in the development of AI reveals a complex ecosystem of risk that includes:

- Risks within the **innovation environment**, where AI solutions are being developed. Risks are occasioned by novelty and the lack of knowledge amongst developers and regulators; and by lack of precedents to resolve ambiguity in regulation concerning automation in AI products and services.

- Risks within the **internal operating environment**, or computational system. Risks are occasioned when well-developed and well-behaved AI models exceed their boundaries; by verification processes that separate software from hardware; and by how AI is connected to and interacts with the component parts of the broader cyber-physical system in which it is embedded-in-use.

- Risks within the **external operating environment**, or human system. Risks are occasioned by user actions and workflow configurations that influence system performance and introduce human errors through complex divisions of labour; and by customer or user requirements that trade-off automation versus accuracy and introduce machine errors.

- Risks within the **regulatory environment**. Risks are occasioned by change management and the cumulative effect of frequent incremental modifications; by a lack of technical expertise to support certification; and by the interplay between regulation and enterprise that directly impacts lead times and development costs.

Seen from the various perspectives of our participants, risk is complex and heterogenous in nature, but if a thread runs through it then it weaves itself around *software*. While well-defined regulatory frameworks and international standards are in place to manage risks in the development of medical devices, for example, these were primarily designed (as regulation in other domains) for *hardware*. The emergence of software both as a component on a device or as a device in its own right (FDA 2018), including software that continues to learn once it is put into service, occasions significant risks.

*4.1.1 Novelty*

Foremost of the risks accompanying AI are those occasioned by *novelty* and the lack of knowledge that accompanies innovation, including those actually involved in developing AI systems.

> Robert: So when you (the safety engineer) put your signature on this - on the basis that you're going to go to court when the accident happens - what was it you thought you were signing? We were looking at somewhere between 18 and 23 sensors in a braking system, every sensor is augmented by machine learning. But when you were looking at somebody signing off the braking system, what we've been through in interviews with the safety engineers that have to put their name onto that was, "There is no machine learning on my vehicle."

Lack of knowledge is the constant companion of innovation. In the above extract, we can see that safety engineers were *unaware* that machine learning was used in the vehicle's ABS, let alone what the effects of an ABS system that uses machine learning might be on other in-vehicle systems or between-vehicle networks. Novelty constantly hampers the apprehension of risk. UK regulation, for example, requires that medical device manufacturers submit an 'intended use statement' classifying the risk status of a device (Class 1 is low, 2 medium, 3 high), which is subject to conformity assessment and certification (GOV.UK 2020). However, such classifications rely on *precedent* to identify and qualify the risks associated with a particular device class and it is not at all clear where automation *sits* within such classification frameworks:

> James: This is the problem: the concept of automation isn't in the classification rules beyond really broad concepts like whether you are informing or driving treatment. So, there's a difference between saying "I think you should do this, you press the button", or "I think you should do this, I'm going to press the button unless you tell me not to". That is the human in the loop/human out of the loop sort of stuff in the AI world, in automation speak, I think.

Little wonder then that the classification of risk in devices that use AI is *ambiguous* in practice and something that often takes time to clarify and establish:

> David: There is an acceleration in manufacturers wanting to have novel AI in their devices. In that case, there is no device against which (it) can be compared against, and they won't have determined an acceptable accuracy so I have to come up with some other method. It's down to the manufacturer to do that. Nonetheless the regulators might have an opinion about whether you have gone about that in a fair way.

The risk status of an AI system cannot always be read off classification frameworks because of the novelty of the system. Instead, the risk status needs to be established through comparisons and performance metrics, subject to regulatory approval.

*4.1.2 Exceeding model boundaries*

The risks of AI are not all about getting to grips with 'unknown unknowns' in the process of innovation, but are also occasioned by a variety of more tangible issues. Both our developer and cyber-security expert recognised the problematic nature of AI models:

> Robert: It is impossible for a digital system to be a complete and accurate representation of your analogue real-world system. It is only a model.

> Julia: Many (design) considerations are no different than if you were using traditional mathematical models and statistics, in that if you use sort of traditional mathematical or statistical techniques to come up with a model you have to be very careful whether you're doing interpolation or extrapolation, because if you know you're going outside the bounds of that model, then there's potentially more risk for not being able to accurately predict something. Those are similar considerations with machine learning conditions and with machine learning algorithms. You know there's an area which you can confidently apply that model to certain problems, but outside certain bounds it's either nonsense or the accuracy just reduces considerably.

There has of course already been a great deal of discussion around the risks that attach to AI models, particularly around 'algorithmic bias' and subsequent 'unfair' discrimination along with the idea that this can be fixed through appropriate data governance (HLEG 2019). However, what is striking here is that even well-developed and well-behaved models are recognised to be potentially problematic insofar as *exceeding model boundaries* affects both the accuracy of AI predictions and limits AI's ability to do prediction in the first place, either of which may introduce further risks. The issue here is not one of fixing a problem in the first instance, but of being aware of and understanding the *limitations* of AI models so as to avoid risks when systems are in use.

*4.1.3 The limitations of software verification (hardware)*

One way in which AI developers have sought to address such concerns is through software verification to demonstrate the 'correctness' of a system. This is extremely challenging for AI systems (Wing & Wooldridge 2022), but even if it were possible there are limits to software verification that directly impact risk management:

> Rene: There are companies that that have no hardware, they sell no hardware. Beforehand people saw software as part of the medical device and often it was embedded inside; it was a bit of software running inside the medical device. But now you see people maybe wanting to have something on iPad that does something and they go "I'm not responsible for the iPad, if you got any issues with that you have to talk to Mr. Apple." From a computer science point of view that makes sense, because if you can run on Apple and you've got another version it can run on Android then you could almost say, well you know the hardware seems not so important. But the FDA and the EU feel very uncomfortable about that. What happens if the Android version suddenly has bugs in it that Apple doesn't, how do you deal with that?

It is inevitably the case that AI models are *embedded* in cyber-physical systems, whether it be the systems they were developed on, or dedicated operating systems or the systems underpinning one of a range of general purpose devices they may be used on, and regulators need to be assured that AI works on them as it should. Software verification is not a panacea to risk in AI systems.

*4.1.4 The complexity of systems (AI \*is\* embedded)*

AI is at root a software component in a broader system, and it may of course be only one software component out of many. The *complexity* of the system AI is embedded in raises tangible risks:

> Robert: These things are all connected in order to get the effect of braking, accelerating, steering. In order to make sure it does the right thing in an ABS system it is extremely important, you know, at an interval somewhere between 100 and 1000 times a second, that you control the coordinated action of a wide area network across your vehicle. There's no good accelerating at a time when somebody else is trying to brake. (But) the control systems for all of these things are often in conflict with each other. I've got privacy obligations in relation to some of this stuff, how I brake and the rest of that stuff is going into my telemetry box. At the same time I'm trying to do a safety critical system. The normal strategy from a security point of view is to encrypt it (but) if you encrypt that data that I need in order to brake then all I have to do is get rid of the (security) key and you can't brake.

> Julia: What we're dealing with is a lot of components that have to act together as a whole. Within the software and hardware there are a huge number of subcomponents. That's an inherent problem of complex systems in general, when we're putting together things that have been developed almost entirely separately and now suddenly, we need to understand how they act together.

> Rene: Yeah, and we also have suppliers of software, right? So, they find an issue, or they have a new version, and don't forget operating systems they become obsolete, so you eventually have to move to the next version and then (that raises the question) does all your stuff still run, including all the drivers on that new version?

As we can see in our participants' shop talk, risk in AI is not only occasioned by the AI component itself, but how it is *connected* to the broader cyber-physical system in which it is embedded-in-use and how the different parts of the system act together as a whole, including those provided by external suppliers. As Robert makes painfully clear with his ABS example, the risks that arise from failing to understanding the *interaction* between AI and other components in complex systems are substantial.

*4.1.5 The impact of the (external) environment*

It is not only interaction within the internal environment, the array of components that make up the of a cyber-physical system, that occasions risks in AI:

> Robert: On a modern vehicle, the high-speed emergency braking comes in through the ABS system. (The system is) essentially looking up ahead and it's going, "Look there's somebody about to step out and we're going to run into them." On those ones actually all I have to do is shine bright lights at the sensors in order to overwhelm them. LiDARs and things of this nature are actually CMOS sensitive areas. All you're doing is stimulating that sensor, you know. Now I don't even have to get into the IT. What I could do is, for every vehicle that's got that, I could turn the brakes on and all I've got to do is flash the headlights.

As Robert again makes painfully clear, risks of AI are also occasioned by the external environment in which it is situated, the *human context* in which cyber-physical systems are placed and operate, where simple actions like flashing headlights (as people routinely do for all manner of reasons) can introduce serious risks, as Rene continues to elaborate:

> Rene: Now all of this has to be seen in how it's being used, and that is where the problem for us often is. If you have a very tightly controlled system, for example in Scotland, (where disease) screening is done by one service, by a small number of people, there's a whole flow control about which work goes to humans, which goes to computers. In England they had a really complicated workflow: first the lowest paid person looks at it, and if they're not very happy, then they send it to an arbitrator and there is a whole flow of different people that get more and more expensive as you go through. So, they put (the automation) in two different places and get different sensitivities and specificities out of it, because overall the system now behaves differently, and it's really complicated to understand what is the risk? What is the impact of (human) error on your system?

> Rene: What we came up with was an AI that classifies the image or says, "I don't know." By doing that we can then start playing with the threshold and say, if we get a bit uncertain, we would rather not (classify the image) and ask the end user. Eventually, we would get 100% accurate: 90% of images automatically done and 10% done by the customer. We ran with that, but some customers were not happy because 1 in 10 (images) they had to do something. That means 1-in-100 will go wrong. Pretty soon you start having quite complicated discussions with people about what is the better way to go with this stuff. Is it more automation, or is it more precision or accuracy? You know, which of the two is preferred. You can't have both at the same time all the time.

While there is a great deal of hype about how AI can improve workflow, Rene makes it visible that *how workflows are configured* directly impacts the performance of AI and may introduce risks of human error in the use of AI systems, specifically occasioned by the separation and distribution of tasks and embedding AI in complicated divisions of labour. Adding to that, are *customer or user requirements* and the trade-off they occasion between automation and accuracy. As Rene makes clear, you can't have both all of the time, and that means there is risk of machine error, of (for example) 1-in-100 going wrong. The numbers aren't what matter (it doesn't matter if its 1-in-10 or 20 or 30, etc.), what matters is the nature of the error introduced by the trade-off: if its classifying eyes as left and right, then not so much of a problem, but if it's classifying potential diabetic indicators or breast cancer or brain tumours then that's a different proposition. Risk here is not confined to the data subject either, but also impacts the user:

> David: We haven't done an economic analysis regarding the total cost to the healthcare system and to patients, (but) if it's a slow progressive disease then a false negative maybe not be as severe as you might expect, because the patient, they're going to come back and maybe get detected. In that case, it might be that actually a false positive is a much higher cost to the system, because then you are referring patients, you are worrying patients, you are causing extra referrals.

Ultimately, as David makes clear, it is not only that the external environment in which AI is used introduces risk, but in doing so a cost is incurred by users and data subjects alike, human and financial.

*4.1.6 Regulatory risks*

Software development and regulation work at very *different paces*, which also introduces risk into the development of AI:

> James: Getting (regulation) to run smoothly, at a pace that the software development and the AI communities are comfortable with – there's a gap there, (it's) not ideal for continuous learning products.

The 'gap' revolves around the speed of change in software development (e.g., frequent updates) versus the speed of regulatory approval (which is glacial in comparison). The gap creates tensions in change management, a situation compounded by AI models that continue to learn after they have been put into service. The proposed solution is to put change management mechanisms in place (e.g., Predetermined Change Control Plans or PCCPs and post-market monitoring). However, it is not clear how such mechanisms will prevent predetermined changes to AI systems (security patches, model updates, usability improvements, etc.) *incrementally* producing changes that taken together would constitute a *substantial* modification requiring regulatory approval.

The gap between software development and regulation also extends to the certification of AI systems:

> James: We have a dispersed system of approved bodies and in EU they're called and 'notified bodies'. These are private companies that hire individuals with technical expertise and regulatory expertise to do that assessment process on behalf of the authority in each country.

> Rene: In principle, they could say "Can I see the source code?", but in practice they will not. I mean they don't have the expertise for that, they just want to see that all the processes around the development of that source code and how it was all stuck together is properly maintained, but they don't want to actually go through the source code and check it, you know.

Checking source code, data training sets and models, test results, etc., is part and parcel of the raft of measures in the AIA (2024) for certifying AI systems as fit for purpose and deployment. Yet we currently lack the necessary expertise to assess novel AI systems at a technical level, which inevitably introduces further risk, there simply aren't enough trained people available to do this yet.

The gap between AI development and regulation also creates risks for the developers when they introduce AI into their products.

> Rene: Always when the regulatory department is at the table the question is, "Are you changing the intention for use, yes or no?" Of course, medical devices change all the time, maybe a supplier goes bust and we have to get a new component in and record that, but we haven't changed the fundamental reason why we use the device. If I make a change to the medical device that changes the fundamental reason for using it, i.e., the intention for use has changed, then we have to go back to the regulator. If the regulator says, "We think that's a substantial change to the device, please go and do a clinical trial and measure all these particular parameters and then come and file £510K with us again", that takes up to one and half years.

Intended use or 'purpose' is also a feature of the AIA. A great many products are now being augmented with AI to offer new and/or improved services that change intended use, even if only slightly. As Rene makes clear, this may not only occasion considerable financial costs (not only filing substantial amounts of money with some regulators such as the FDA and having to pay for clinical trials, but also establishing and maintaining a quality management system for certification), it also significantly impacts time to market. AI thus creates significant *market risks* for developers in the necessary interplay between product development and regulatory compliance and certification.

**4.2 How do you manage risk?**

We do not claim this is a comprehensive or exhausting register of risk in anyway, only that it surfaces iterable epistopics of risk in AI. What we want to do next is elaborate something of what participants know about managing these risks.

*4.2.1 Managing risks in the innovation environment*

The risks occasioned by lack of knowledge, ambiguity and the lack of precedence have thus far been addressed by our participants in two distinct ways: 1) by setting performance benchmarks and 2) by attacking an AI system to surface the risks it presents.

> Rene: What you sometimes have to do is bring a lot of well-known experts to the table and say these people have said this is the minimum accuracy you'd need for a viable system. If you are able to persuade the regulator that this is the right team to make that call, and your technology could actually meet it, then you're good to go.

Persuading the regulator is not only a matter of doing in-house tests to determine performance metrics and having experts confirm them. Human benchmarks may be set by professional bodies and AI systems are required to perform at least as well where novel systems are concerned and expected to perform better with respect to modified systems. However, it is also the case the far more stringent measures may be required:

> James: If you've done all these studies in-house to figure out the answers to the safety questions, but ultimately your notified body feels that you can't deploy that product safely without doing some *proper user testing* in the real world, then you need to do something called a clinical investigation. The protocol, the technical information, the risk assessment information, regardless of class, all of that comes to MHRA for approval.

The complement to benchmarking (as ironic as this may sound) is to systematically attack an AI system through adversarial machine learning or adversarial testing as it is also called. The aim isn't to test if an AI system works as it should (functional testing), but rather to see if and how it can be subverted. Adversarial testing thus seeks to both 'attack' and 'poison' AI systems directly and to manipulate variables in their operating environment to foster understanding of the risks created by novel systems:

> Robert: You're designing control systems based around the presumption in the margins of errors that will be acceptable but actually, in general, I found that as soon as you do any poisoning attacks or you change the environment in anyway, its quite easy to persuade them to run outside that margin of error.

Adversarial testing is a critical part of the mix in managing the risks presented by AI systems. Methodologies are emerging to support adversarial testing at scale (e.g., ResiCAV 2024) and it is key to enabling compliance with emerging regulation. The AIA (2024), for example, requires the "estimation and evaluation of the risks that may emerge" when an AI system is used "under conditions of reasonably foreseeable misuse" (Article 9). It is hard to see how the risks accompanying reasonably foreseeable misuse are to be apprehended without widespread, systematic uptake of adversarial testing in AI development. There is a sense in which adversarial testing speaks to the "demonic order" of objects, "hidden orders of contingency" not usually encountered in the ordinary course of action and interaction with them (Garfinkel 2022). Unusually, and distinctively, adversarial testing is engaged in *surfacing* the demonic order and making it available to mundane practice for remedial action.

*4.2.2 Managing risks in the internal operating environment*

Our participants manage the risks occasioned by the limitations of AI models through good machine learning practice:

> Rene: As we heard during the pandemic, pulsometers that measure oxygen in the blood were not as accurate on people with darker skin than those with fairer skin. If you start with the wrong data, it will just follow what you're doing, it doesn't do anything else than that. It doesn't think outside of that space that you've given it in terms of samples.

> David: One of the basic principles is that your test set is separate from your training set when you evaluate performance, but also that it should be independent, the Good Machine Learning Practices published by the FDA (FDA et al. 2021) says that. But we know that, for example, there are differences between the way that our customers use devices, and there are artefacts that can come into play that were not present in the training set - unwanted details on the images, such as dust, or people not being able to hold still, sometimes body parts can get in the way.

> Rene: Does it suddenly think it's a disease or does it obscure the disease? If you are not aware of that, it just keeps learning. So, *understanding what goes in* is important.

As Rene and David make perspicuous, good machine learning practice isn't just a matter of good data governance, of sampling, training and testing (etc.), and following regulatorily guidance. It does include these things, of course, but critically there is more to good practice that turns upon understanding the environment in which an AI system is placed, understanding how it is used, understanding how the environment effects it, how dust, and people not being able to hold still or body parts getting in the way (etc.) may impact the ongoing process of machine learning.

Understanding what goes in, not just at the beginning but over its lifetime, and understanding how that impacts or may impact the AI model is key to ensuring it does not exceed its boundaries.

The risks occasioned by software verification and its separation from hardware are dealt with by restricting access to specific hardware platforms:

> Rene: There are small companies that are building AI for medical devices who say, "Give me these images, I'm going to build this machine learning and it will just run everywhere." The first one that tried, they don't have camera, right? They are basically depending on other vendors to provide those images. The FDA went, "What did you test it on?" "Well, we use this model of this particular vendor" "Well, OK, we will grant you permission to sell this, but *only* for that model of that particular vendor."

Similarly, the MHRA requires the whole system to be safe – not only the software – and that the developers of stand-alone software "demonstrate compatibility" with recommended hardware platforms (MHRA 2022).

Risks arising from an AI system's connection and interaction with the broader cyber-physical system in which it is embedded-in-use are addressed through testing, functional and adversarial (and no doubt a raft of other tests too). However, this is an *extremely challenging* area:

> Robert: The assumption is that you can do empirical testing and get a conclusion out of that, but the issue you've got is *complexity that you can't measure*. (In) the interplay of the digital and the physical there seems to be an assumption that if you lock down the digital part then our systems will be more resilient and more reliable, but actually, the environment that they're working in, because they're all *connected* to each other, that isn't true. So (for example) I stimulate the machine learning so that the control system said to the brakes, "We don't know what state they're in, we're going to have to stop the car." OK, so that sounds like a very sensible thing to do. However, if over a one-hour period that happened to 40 million vehicles —— most things we've needed to trust in the digital world have been extremely small-scale systems. That doesn't really apply to robotics. It doesn't apply to smart buildings. It doesn't apply to the auto industry. It doesn't apply to any of the major volume areas, and it certainly doesn't apply to AI. I think there is a step which is you as an engineer, given that there is this gap in knowledge, have to resolve to your satisfaction, such that were you ever called to court, you have an answer. I honestly don't know how to do that at the moment. I mean, honestly, I don't know how to do that.

While it may well be tractable to manage the risks of AI embedded in small-scale, stand-alone products through benchmarking, testing, platform restriction and understanding the impact of use on AI models, and while this may cover a great many actual uses of AI, the complexity of an envisioned near-future world populated by networks of autonomous systems is an entirely different proposition. As Robert makes visible, it is not at all clear how we can 'measure' and otherwise apprehend the complexity of, and with it the risks arising in, the massively connected environment in which it is hoped that AI will soon be embedded-in-use. A sobering thought, especially as it comes from a cyber-security expert in a very large international electronics company, and one that makes it perspicuous that risk is *not always* calculable. As Beck (1986) observed,

> "under the surface of risk calculation new kinds of industrialized, decision-produced incalculabilities and threats are spreading within the globalization of high risk industries … Along with the growing capacity of technical options grows the incalculability of their consequences …  in the risk society the unknown and unintended consequences come to be a dominant force in history and society."

Risk cannot always be calculated, hence Becks pessimism about risk management frameworks. The globalisation of high industries such as AI introduces "demonically wild contingencies" (Garfinkel 2022), which may not necessarily be tamed by adversarial testing. It may, however, at least make us cognisant of significant risks not registered in risk management frameworks.

*4.2.3 Managing risks in the external operating environment*

Adversarial testing also plays a role in understanding environmental effects and how the actions of users might impact AI. Our participants try to address potential costs to data subjects and users through quality management systems (ISO 9000 2015) and by putting measures in place to prevent or mitigate risks in the external operating environment.

> Rene: There's logic built into the system that says "OK, I've now determined this device is no longer safe." It will then start switching things off, right. It will switch off the power. It will switch off the laser. It will put a shutter down. Our laser safety system is not software engineered. It's hardware engineered. It runs on its own electronic board and cannot be accessed from software. You cannot tamper with that. The only way to tamper with it is to open the hatch, and if you do that then the tamper mechanism blocks the whole device. Everything is driven by risk. *Risk is everywhere.* Whether that is the risk of electrocuting a patient or the risk of having radiation emitted by your device because they are wearing a pacemaker. You must have measures in place to mitigate against those risks. At the end of the process the device is going to an independent company that measures how much radiation there is to check that it's below a standard that can be used on people with pacemakers and so forth. All that has to be built into your processes and be recorded into a (quality management) system that is trusted by the regulator.

Risk is everywhere. However, the risks around hardware and how it might impact and be impacted by the external operating environment are much better understood than with software, governed by all manner of international standards and engineering best practice. Much as the embedded nature of AI represents a huge challenge, at least at scale, then so too the ways in which AI is impacted by and impacts its external operating environment. There is, particularly, a huge gap in our understanding of how workflow configuration and customer requirements impact and are impacted by AI and the risks that emerge in practice. Furthermore, these are issues that necessarily stand *outside* quality management systems, as they turn on what customers and users do with AI systems after products and services have been designed and been put into service. While good machine learning practice would have developers understand something of what this entails and how it affects AI, how that should be done and done at scale is an open question. The regulatory notion of 'post-market monitoring' sounds as though it would catch this, but when one looks it can be seen that it does not focus on these issues but rather on compliance (see Articles 12 & 61 in the AIA, for example). It may, in theory at least, be perfectly possible to have a compliant system whose performance is nevertheless adversely affected by workflow configurations and user requirements. The focus of regulation is on the effects, but understanding *cause* is equally important.

*4.2.4 Managing risks in the regulatory environment*

The risks occasioned by change management are currently managed by restricting machine learning to *pre-market* development:

Julia: The machine learning that at the moment might be approved for medical devices is something that has been trained and then 'cemented' – it will not be allowed to change how to solve the same problem in a different way. The essence of machine learning as it came about originally is that it continues to learn, it continues to come up with new solutions for old problems, or else the problem changes and it's allowed to adapt. That's how it's continues to be applied in other areas. Perhaps we will not do that in a medical device because the algorithm will continue to change itself and produce new solutions, but there is no human intervention, we cannot check what's happening or try to understand why it changed its solution when presented with a new problem, and that of course is a huge, huge, potential risk. So we're curtailing machine learning at a very, very definite point and stopping it.

In addition to 'cementing' machine learning at a definite point, which is then subject to regulatory approval, we have also seen that regulators constrain AI to particular vendor devices and platforms to manage risk. However, the landscape is changing, and regulation is opening up to AI systems that present significant risk (e.g., Class 2 or 1 devices) and continue to learn after being placed on the market. The FDA, Health Canada and MHRA have jointly laid down principles for the use of machine learning in medical devices, including that they be monitored for "retraining risks" occasioned by overfitting, unintended bias, dataset drift and any other forms of model degradation once put into service (FDA et al. 2021). The AIA (2024) similarly envisages "high-risk AI systems that continue to learn" being placed on the market and that predetermined changes "shall not constitute a substantial modification" requiring further regulatory approval (Article 43).

Nonetheless, the cumulative effect of frequent incremental modifications remains problematic, as does the technical expertise required for compliance assessment and certification:

Rene: Perhaps in the future there may be a machine learning company that that doesn't actually build machine learning but has the expertise to say, "I know how to at least check that if somebody has made machine learning, they've done it properly, here's a certificate."

The AIA, which like GDPR will have global reach, envisages that notified bodies will have the expertise to assess source code if it is needed to determine the compliance of an AI system (Article 64). However, as Rene pointed out above, the expertise is not yet available. There are, then, significant risks built into the regulatory system, though just as GDPR let to exponential growth in Data Protection Officers, so too we might expect exponential growth in machine learning assessment along the lines Rene envisages.

There is perhaps more tangible progress towards managing market risks through regulatory sandboxes to help companies innovate in safe ways (see, for example, ICO 2020), though again there is a long way to go in terms of fostering innovation:

> Rene: It's regulated context really. I mean we're working on all sorts of AI and there's another meeting next week that's just popped into my calendar about how far are we're going to push this towards the regulator. Already some people in the company are getting very scared about it, you know, "The regulator might ask us to do all these really expensive things." So yes, it's the regulator really, keeping us under control. I think if that wasn't there, there would be quite creative solutions out there.

As we can see from what our participants know about managing risk in AI, it is far from a done deal. Some risks are tractable, others present significant challenges. The impact of complex large scale networked systems, the human environment on AI, incremental modifications and change management, and slow-paced regulatory systems on innovation and economic development are notable.

## 5. DISCUSSION

Our brief study of the epistopics of risk known by three members of staff involved in the manufacture of medical devices that use machine learning, by one cyber-security expert assessing machine learning in vehicles for a large international electronics company, and one policy expert engaged in the regulation of medical devices in the UK has surfaced a range of risks in the innovation environment, the internal and external operating environment, and the regulatory system. These epistopics make it perspicuous that the novelty of AI is seen and treated in practice as something that is inherently risky and might be reckoned with through benchmarking against human performance; that steps need to be taken to apprehend reasonably foreseeable misuse of AI and to understand what goes into an AI system in practice (in contrast to training) and how this affects AI over time, including how and why an AI system continues to change during post-market deployment; that there is need for developers to take steps to understand how AI connects to and interacts with the broader cyber-physical system in which it is embedded-in-use, and how the complexity of the cyber-physical system impacts AI; critical too is the need to understand how AI is configured and situated in the external human environment and the impact this has on AI; to develop the expertise needed to assess and certify AI systems; and to figure out how regulation might itself become part of the innovation pipeline to mitigate potential economic risks.

The epistopics of risk identified in our study are not psychological phenomenon in-the-head of our participants, not merely the opinion or point of view of five random people, but are known as part of their own *organised enterprises*. They are, to use Garfinkel's turn of phrase, 'magnificent methodological things' that order AI development in practice. They are *organisational objects*, situated features of the workplace, and it is because they are organisational (not psychological) objects that we think they generalise and reach beyond our five participants and the three organisations where they work. Simply put, it is impossible to imagine our participants are the only ones who have to contend with the risks posed by the innovation environment, the internal and external operating environment, and the regulatory system. Furthermore, that others involved in the development of AI have to contend with these epistopics of risks as they are known within their own organised enterprises underscores that they are iterable, and that they are iterable opens up the space for further discovery. Indeed, we think the iterable nature of epistopics of risk creates the potential for an interdisciplinary program of inquiry having at least three distinct functions.

### 5.1 Iterable epistopics of risk and AI research

Iterable risks in the innovation environment, the internal and external operating environment, and the regulatory system may provide a focus for interdisciplinary research informing the ongoing development of AI. How, for example, might AI performance be benchmarked and precedents be established? In what ways does the external environment impact AI? How can we verify the relationship between software and hardware? How can we measure complexity and manage the impact cyber-physical systems have on AI? In what ways do customer requirements and workflow configurations impact AI? How do we manage change? How can we develop the relationship between regulation and AI development to support innovation? We are aware that researchers have begun to explore the epistopics of risk. Legal-tech scholars have explored novel ways in which regulation might foster innovation (Luger et al. 2015), for example. Design fiction has been leveraged to explore AI in future environments (Sailaja et al. 2019, Pilling et al. 2022) and exposed risks to privacy and fundamental rights. Field studies and ethnography has been used to understand the impact of customer requirements and workflow on AI (Sendak et al. 2020, Bede et al. 2020). These are but a few examples. Nonetheless, they underscore the point that the epistopics of risk are not only iterable but in being so provide a resource that might drive an interdisciplinary program of inquiry into risk and risk management in AI.

**5.2 Iterable epistopics of risk and AI design and development**

A second contribution focuses on supporting the design and development of AI, particularly with respect to regulatory compliance. The AIA (2024), which like GDPR has global reach, mandates that steps be taken to eliminate or reduce risks (Article 9) and this includes giving 'due consideration to the user and the environment' in which a system is intended to be used. Nothing in the regulation says how this might be done. However, HCI provides human-centred design methods could be used to elicit epistopics as users know them in their own organised enterprise. From its early concern with ergonomics and engineering computing systems around human factors (Benedyk 2020), to the design of the software control dialogue to support human-computer interaction (Card et al. 1983), to the turn to the social and shaping technology around organisation (Grudin 1990), users and their environment have been at the heart of HCI. This core interest is accompanied by a battery of research methods for understanding users and their environment and for actively involving them in design and development too: design ethnography, contextual design, scenario-based design, participatory design, etc. (IDF 2024) could easily be purposed to enable the iteration and discovery of epistopics of risk during the development process. In doing so, they might help AI developers comply with the 'human-machine interface' requirements of the AIA (Article 14), which go beyond interpretation and 'explainable AI' (Abdul et al. 2018) to mandate the provision of tools enabling users to *intervene and take action* to prevent or minimise risks to health, safety or fundamental rights (Urquhart et al. 2022).

**5.4 Iterable epistopics of risk and AI governance**

A final contribution, at least at this point in the evolution of epistopics, focuses on supporting the governance of AI. We began this paper with a brief discussion of general calculus: risk management frameworks that provide a method or procedure for reasoning about risk and rendering it accountable and thus tractable, such as algorithmic impact assessments, ethical principles, international standards, and regulations. To the best of our knowledge, *none* of the epistopics of risk identified in this paper are encapsulated by general calculus. They are absent from general calculus. Missing. They are missing because general calculus trades in what is known, established risk, whereas the epistopics uncovered by our study are very much at the forefront of innovation: the risks occasioned by novelty and lack of knowledge, or the limitations of AI models and software verification, or the impact of complexity and the environment, for example. Epistopics of risk, as known within the organised enterprise and situated practices of those actively involved in AI development, may thus sensitise general calculus to its *blind spots*. The annexes to the AIA (2019), for example, show that EU clearly expects AI to be embedded in a complex array of machinery. Yet consideration of the scale of complexity – of the *massively connected* cyber-physical systems including the transport networks it is envisaged AI will be embedded in – is entirely absent. It doesn't figure in legal reasoning about the risks of AI, though situated practice suggests that it may well pose the most significant risk of all as it is currently immeasurable. The epistopics program may prove a useful complement to the evolution of general calculus.

**6. CONCLUSION**

It is broadly acknowledged that trust is key to the uptake and success of AI in society. For many involved in the development of AI – which includes our political and legal representatives and industry drivers as a well as computer scientists and engineers, philosophers and social scientists – trust turns in significant ways upon risk management and the avoidance or serious limitation of harm. Risk management frameworks consequently abound, providing general calculus or methods of reasoning about risk and how to deal with it. However, general calculus is necessary but not sufficient for the management of risk in AI. It trades in established risk, not risk as it is known at the forefront of innovation in the organised enterprises and situated practices of those who are actively involved in doing AI development. We propose an epistopics program to complement general calculus, support the design and development of AI systems, and drive interdisciplinary iteration and continued discovery of epistopics of risk in AI research. The initial contribution, provided by a short study of five people in three organisations involved in AI development, nevertheless surfaces epistopics of risk that have reach and apply in other settings. They highlight risks in the innovation environment, the internal and external operating environment, and the regulatory system. They are iterable. They may be further investigated in the organised enterprise of others involved in AI development to elaborate how risk is known and managed in practice, and trust is thereby built into AI, as well the significant challenges that lie ahead.

## 7. STATEMENTS AND DECLARATIONS

**Competing interests statement:** The authors have no financial or non-financial interests that are directly or indirectly related to the work submitted for publication to disclose.

**Data availability statement.** The data on which this paper is based is subject to non-disclosure agreement and is not available to the public.